# AMINO ACIDS IN COMETS AND METEORITES: STABILITY UNDER GAMMA RADIATION AND PRESERVATION OF CHIRALITY

Susana Iglesias-Groth[1,2], Franco Cataldo[3,4], Ornella Ursini[5], Arturo Manchado[1,2,6]

[1]Instituto de Astrofísica de Canarias (IAC), Via Lactea s/n, E-38200 La Laguna, Tenerife, Spain.
[2]Departamento de Astrofísica de la Universidad de La Laguna, Spain
[3]Istituto Nazionale di Astrofisica. Osservatorio Astrofisico di Catania, Via S. Sofia 78, 95123 Catania, Italy.
[4]Actinium Chemical Research, Via Casilina 1626/A, 00133 Rome, Italy.
[5]Istituto di Metodologie Chimiche, CNR, Via Salaria Km 29,300, 00016 Monterotondo Stazione, Rome, Italy.
[6]CSIC, Spain.

**Abstract**

Amino acids in solar system bodies may have played a key role in the chemistry that led to the origin of life on Earth. We present laboratory studies testing the stability of amino acids against γ radiation photolysis. All the 20 chiral amino acids in the *levo* form used in the proteins of the current terrestrial biochemistry have been irradiated in the solid state with γ radiation to a dose of 3.2 MGy which is the dose equivalent to that derived by radionuclide decay in comets and asteroids in $1.05 \times 10^9$ years. For each amino acid the radiolysis degree and the radioracemization degree was measured by differential scanning calorimetry (DSC) and by optical rotatory dispersion (ORD) spectroscopy. From these measurements a radiolysis rate constant $k_{dsc}$ and a radioracemization rate constant $k_{rac}$ have been determined for each amino acid and extrapolated to a dose of 14 MGy which corresponds to the expected total dose delivered by the natural radionuclides decay to all the organic molecules present in comets and asteroids in $4.6 \times 10^9$ years, the age of the Solar System.

It is shown that all the amino acids studied can survive a radiation dose of 14 MGy in significant quantity although part of them are lost in radiolytic processes. Similarly, also the radioracemization process accompanying the radiolysis does not extinguish the chirality. The knowledge of the radiolysis and radioracemization rate constants may permit the calculation of the original concentration of the amino acids at the times of the formation of the Solar System starting from the concentration found today in carbonaceous chondrites. For some amino acids the concentration in the presolar nebula could have been up to 6 times higher than currently observed in meteorites. It is also expected the preservation of an original chiral excess. This study adds experimental support to the suggestion that amino acids were formed in the interstellar medium and in chiral excess and then were incorporated in comets and asteroids at the epoch of the Solar System formation.

# 1. Introduction

One of the breakthroughs in astrochemistry and astrobiology occurred in recent years was the unequivocal discovery of chirality in carbonaceous chondrites (CC), the rather rare meteorites which contain a relatively high level of macromolecular carbon commonly known as kerogen (Cronin and Pizzarello, 1997; Pizzarello and Cronin 2000; Pizzarello et al. 2008). More in detail, the enantiomeric excess was measured in the amino acids fraction extracted from the CC. The absolute certainty that the measured enantiomeric excess does not derive from contaminations is due to the fact that not only common α-amino acids used in the biochemistry of living organisms have been found in the extracts from CC but even really uncommon β-amino acids (Pizzarello and Cronin, 2000; Pizzarello et al. 2008) and diaminoacids (Meierhenrich et al. 2004; Meierhenrich, 2008). This fact excludes any contamination from terrestrial microorganisms and confirms the abiotic origin of the observed chiral excess. Furthermore also the isotopic signature confirms the non-terrestrial origin of the organic molecules extracted from the CC (Engel and Macko, 1997). As a further proof of the abiotic origin of the amino acids present in meteorites it is reported that the distribution of the abundances of the different amino acids follows that observed in the laboratory abiotic synthesis according to the Miller (1953,1955 and 2000) or the Sagan and Khare (1971) derived reactions. The abiotic origin of chirality is hence an experimental fact, and chirality is a key property of the organic molecules involved in the biochemistry of life. Without chirality there is no life because all the chemical reactions occurring in living organism are asymmetric synthesis and cannot occur or cannot be as fast and as selective as they are thank to chirality of the molecules and macromolecules involved (Lough and Weiner, 2002; Cataldo, 2007a).

There are several hypotheses on the mechanism which has led to the experimentally observed chiral excess. The most convincing mechanism involves the action of circularly polarized light (CPL) on racemic mixtures of molecules formed in the interstellar medium before the formation of the Solar System (Meierhenrich, 2008). Because of the slightly different light absorption cross section in a racemic mixture one enantiomer absorbs a slightly higher amount of CPL than the other enantiomer. A prolonged irradiation of a racemic mixture of molecules with CPL transform it into a scalemic mixture, hence with a chiral excess. The effect of UV-CPL has already been demonstrated experimentally in creating a chiral excess (Jorissen and Cerf, 2002; Meierhenrich et al. 2005; Meierhenrich 2008). Thus, the scalemic mixture of molecules once formed in the interstellar medium where then incorporated in the primitive Solar nebula and can be found still today in meteorites and presumably other bodies of the Solar System which have experienced only a limited processing, in other words where the matter was preserved as it was $4.6 \times 10^9$ year ago (Sephton 2002 ; Kwok 2009). The point is that there are already known sources of circularly polarized light in certain star-forming regions (Bailey et al. 1998) which have a clear implication in the induction of chirality in the racemic mixture of molecules formed with abiotic processes in the interstellar medium.

To further underline the complexity of the kerogen in the CC, it should be remembered that only 10-20% of the carbonaceous matter is water or solvent extractable, the remaining fraction is composed by a complex mixture of high molecular weight crosslinked organic matter resembling for certain instances the coal of terrestrial origin. Refined analytical techniques have been used to study also this insoluble fraction revealing that it is a really complex mixture of thousands of different molecular species (Schmitt-Kopplin 2010, Sephton, 2002), a complexity comparable or even more impressive than that of terrestrial coal.

The top down approach in the search of the abiotic synthesis of amino acids and the relative chiral excess has involved the analysis of meteorites and in particular of the CC as just reported. In the present paper we address the problem of amino acid preservation in CC and the preservation of chirality.

For the purpose of the present work we have selected 20 chiral α-amino acids, the common amino acids used in the biochemistry of the current terrestrial living organisms with the exclusion of glycine (the simplest but achiral α-amino acid). The 20 chiral amino acids selected were all in the *l*-enantiomeric form, since it is known that the current biochemistry of living organisms is using almost exclusively the *l*-enantiomers. The individual amino acids molecules were exposed to γ ray irradiation in closed vials to a total dose of 3.2 MGy (1MGy= $10^6$ gray) in an approach similar to that used by us in the study of the radiation stability of $C_{60}$ and $C_{70}$ fullerenes (Cataldo, Strazzulla and Iglesias-Groth, 2009).

Comets, asteroids and other bodies of the Solar System, especially those far away from the Sun the Kuiper belt or in the Oort cloud are thought to include very primitive organic material embedded in ices and rocks. Such material has not been processed so much since the times of the Solar System formation till the present and remained embedded inside the host bodies. The almost unique source of degradation and processing of the organic material occurring in $4.6 \times 10^9$ years (the Solar System age) at a depth >20 m in comets or asteroid derives almost exclusively from the radionuclide decay (Draganic et al. 1993).

The Nobel laureate Harold C. Urey (1955, 1956) calculated the amount of energy generated by the decay of radionuclides in comets, asteroids, meteorites and larger bodies of the solar system. More recently, Draganic et al. (1993) reported results of such calculations essentially showing that in comets or asteroids at a depth of 20 meters or more, the cosmic rays are completely shielded and the produced radiation derives almost exclusively from the radionuclide decay. In a time scale of the age of the Solar System, i.e. $4.6 \times 10^9$ years (Unsold and Baschek, 2002), the total radiation produced by radionuclide decay in bulk comets, asteroids and larger bodies of the Solar System is ≈14 MGy. The short lived radionuclide $^{26}$Al is able to produce ≈11 MGy in the first billion years of existence of the Solar System (Kohman, 1997) while the remaining contribution to the total radiation dose of 14 MGy derives from the decay of radionuclide having long and very long decay time like for instance $^{40}$K, $^{232}$Th, $^{235}$U, $^{238}$U (Urey, 1955 and 1956; Draganic et al. 1993). The calculations of Urey (1955, 1956) and Draganic et al. (1993) are in agreement also with other calculations made by Prialnik et al. (1987).

In the present work we report about the decomposition and radioracemization of a series of selected amino acids which have been irradiated in the solid state at 3.2 MGy, a dose equivalent to that received by the aminoacids in comets or asteroids in $1.05 \times 10^9$ years. From the degree of decomposition and radioracemization reached we were able to establish both the relative resistance to radiation of each amino acid under study and to extrapolate the initial amount at the epoch of the Solar System formation.

## 2. Experimental

*2.1 Materials and equipment*

A set of 20 proteinaceous amino acids shown in Scheme 1 (l-alanine [Ala], l-arginine [Arg], l-asparagine [Asn], l-aspartic acid [Asp], l-cysteine [Cys], l-cystine, l-glutamic acid [Gln], l-glutamine [Glu], l-histidine [His], l-isoleucine [Ile], l-leucine [Leu], l-lysine [Lys], l-methionine [Met], l-phenyalanine [Phe], l-proline [Pro], l-serine [Ser], l-threonine [Thr], l-tryptophan [Trp], l-tyrosine [Tyr], l-valine [Val]) were obtained from Sigma and used as received.
The measurement of the degree of decomposition of the amino acids under study was made by two techniques: the Differential Scanning Calorimetric (DSC) analysis which was made on a DSC-1 Star System from Mettler-Toledo and the Optical Rotatory Dispersion spectroscopy performed on a Jasco P-2000 spectropolarimeter with a dedicated monochromator.

*2.2 Irradiation procedure with γ rays*

The irradiation of the 20 amino acids with γ rays was made in a Gammacel from Atomic Energy of Canada at a dose rate of 1.7 kGy/h. A total dose of 3200 kGy = 3.2 MGy (1Gy = 1 J/kg) was delivered to each sample. The samples of each individual amino acids (300 mg each) were irradiated in the solid state insight tightly closed glass vials.

*2.3 Analysis with differential scanning calorimetry*

The irradiated samples were tested for purity by a DSC (Differential Scanning Calorimetry) at a heating rate of 10°C/min in static air. As reference the DSC test was applied also on pristine, not irradiated samples under the same conditions. The amount % of residual sample after the solid state radiolysis $N_\gamma$ was determined from the ratio of the melting enthalpy after the radiolysis at 3.2 MGy ($\Delta H_\gamma$) and the enthalpy before radiolysis measured on the pristine sample ($\Delta H_0$):

$N_\gamma = 100 \, [\Delta H_\gamma / \Delta H_0]$ (1)

*2.4 Analysis of the radioracemization degree by Optical Rotatory Dispersion spectroscopy.*

The degree of radioracemization was measured by ORD (Optical Rotatory Dispersion) spectroscopy. The irradiated amino acid samples were dissolved in HCl 1M and the optical rotation was measured on the resulting solution using a polarimetric cell of 0.5 dm length in the range between 400-600 nm. As reference, the same measurement was also made on standard pristine amino acid dissolved in the same medium at the same concentration. From the ratio of the specific optical rotation after radiolysis $[\alpha]_\gamma$ and before radiolysis $[\alpha]_0$ the residual optical activity $R_\gamma$ has been determined:

$R_\gamma = 100 \, [\alpha]_\gamma / [\alpha]_0$ (eq. 2)

The value of $R_\gamma$ was always averaged in the entire range of wavelengths (400-600 nm) where the specific optical rotation was measured.

## 3. Results and discussion

3.1 – The amino acids stability to radiation and radioracemization: the bottom-up approach

The abiotic synthesis of amino acids starting from simple molecules like methane, ammonia, water occurs in different conditions when there is a sufficient source of energy from electrical discharges (Miller, 1953; 1955; 2000) to UV irradiation (Sagan and Khare 1971). The vacuum UV irradiation of interstellar ice analogs produces a mixture of amino acids and can be considered an extension of the Sagan synthesis (Bernstein et al. 2002; Nuevo et al. 2008). The irradiation with circularly polarized light of interstellar ices analogs produces the asymmetric synthesis of a mixture of amino acids (Muñoz Caro et al. 2002; Takano et al. 2007). Since there are known sources of circularly polarized light in the star forming regions (Bailey et al. 1998), it is thought that molecules like amino acids which are formed in racemic mixtures in prebiotic processes are transformed into scalemic mixtures (mixtures containing an enantiomeric excess) under the action of certain natural circularly polarized light sources. Then, these molecules are incorporated in comets and asteroids and other bodies of the solar system. Notably, as already mentioned, a series of amino acids some of them in chiral excess have been found in different meteorites (Pizzarello and Cronin, 2000; Sephton, 2002; Pizzarello et al. 2008). If adequately shielded from cosmic rays, the organic molecules present in asteroids and meteorites undergo radiolytic decomposition due to the decay of the radionuclides present in the rocks of those bodies. Thus, the questions we intend to answer are: what was the amount of amino acids present $4.6 \times 10^9$ years ago in the bodies just formed at the beginning of the Solar System? By determining the radiolysis half life period we will show that it is possible to answer to that question. Similarly can the enantiomeric excess present in the organic matter embedded in comets or asteroids (adequately shielded from cosmic rays) survive a continuous high energy irradiation for $4.6 \times 10^9$ years? It is well known that the high energy radiation plays always against the preservation of chirality (Cataldo, 2007b; Cataldo et al., 2008) because of the well known phenomenon of radioracemization. Only in exceptional cases the radiolysis of certain substrates may surprisingly lead to an enhancement of the chirality rather than to the radioracemization (Cataldo et al. 2007; 2009).

Accepting the Urey (1955; 1956) and Draganic et al. (1991) calculation that the total radiation dose delivered in $4.6 \times 10^9$ years to the amino acids buried at a depth of >20 m in comets, asteroids and other bodies of the Solar System is estimated to be 14 MGy. From this, it can be deduced that the radiation dose used in our study: 3.2 MGy, corresponds to the dose equivalent to $1.05 \times 10^9$ years. It must be emphasized a limitation between of our experimental approach in comparison to real irradiation conditions derived from the radionuclide decay. In our case a dose rate of 1.7kGy/h was used so that we were able to reach the total dose of 3.2 MGy in 2.6 months. In this relatively short interval of time the dose rate remains constant. On the other hand in comets, asteroids and other bodies of the Solar System, the dose rate derived from the radionuclide decay is much lower and additionally it is not linear with time. For instance, it is well known that $^{26}$Al is completely consumed in less than $10^7$ years at the very beginning of the story of the Solar System. In that relatively short time $^{26}$Al decaying by either of the modes beta-plus or electron capture, both resulting in the stable nuclide $^{26}$Mg, produces about 11 MGy of radiation dose (Draganic et al. 1993). All the other radionuclides ($^{40}$K, $^{232}$Th, $^{235}$U, $^{238}$U, $^{10}$Be, $^{129}$I, $^{237}$Np, $^{244}$Pu, $^{247}$Cm) produce about 3 MGy in 4.6x10$^9$ years. Thus, in the real cometary conditions, the dose rate was not linear at all, it was high at the beginning of the Solar System especially when there was the contribution derived from $^{26}$Al and was slowed down progressively by reaching the present time. In spite of these important differences in the dose rate between the experiment and the real conditions, in radiation chemistry the radiolysis effects are essentially governed by the total radiation dose (Hughes, 1973; Woods and Pikaev, 1994) and in this case the dose we have administered is 3.2 MGy. Of course we have to assume a linear relationship between the radiation dose and time. It is an artificial assumption which simplify our calculations but the final effects on the organic substrate are the same at the end. Thus, if the total radiation dose delivered by the radionuclides decay in 4.6x10$^9$ years is 14 MGy, then 3.2 MGy correspond to the dose delivered in 1.05x10$^9$ years. Once it has been determined the degree of decomposition of a given amino acid at 3.2 MGy, it can be extrapolated to 14 MGy, hence at 4.6x10$^9$ years. Assuming a first order kinetics in the solid state radiolysis of the amino acids we may calculate both the rate constant k and the half-life period T$_{1/2}$ according to the following equations (Yeremin, 1979):

$$N_\gamma = N_0 \, e^{-kt} \qquad (3)$$

$$\ln (N_0 / N_\gamma) = k_{dsc} t \qquad (4)$$

$$k_{dsc} = [\ln (N_0 / N_\gamma)] \, t^{-1} \qquad (5)$$

$$T_{1/2dsc} = (\ln 2)/k = 0.963 \, k^{-1} \qquad (6)$$

Where $N_\gamma$ is the amino acid concentration at any time of radiolysis, $N_0$ the concentration at the beginning of the radiolysis. Since $N_0 = 100$ i.e. a given amino acid it is assumed 100% pure and $N_\gamma$ the residual amount of a given amino acid as measured by DSC at 3.2 MGy, i.e. at 1.05x10$^9$ years, and as determined by eq. 1, it is possible to derive the value of k, the radiolysis rate constant from eq. 5. Once k is known, then it is possible to calculate the radiolysis half life of the amino acid.

The amino acids under study were subjected also to Optical Rotatory Dispersion (ORD) spectroscopy with the purpose to study the radioracemization degree. To avoid artificially induced racemization, the pristine and solid state irradiated amino acids were dissolved in acidic or neutral solution (Djerassi, 1960; Jirgensons, 1973) before the ORD measurement. Thus, the pristine and the irradiated amino acid samples were dissolved in water or in HCl 1M at the same concentration and the ORD curve was measured with a spectropolarimeter. One thing is the radiolysis of a given amino acid without caring about its optical activity and its enantiomeric excess and another thing is instead the evaluation of the residual optical activity after the radiolysis at 3.2 MGy. By using the residual optical activity $R_\gamma$ derived from the ORD curve, and making the same mathematical treatment adopted for $N_\gamma$, it is possible to derive the analogous equations:

$$R_\gamma = R_0\ e^{-kt} \tag{7}$$

$$\ln(R_0/R_\gamma) = k_{rac} t \tag{8}$$

$$T_{1/2rac} = (\ln 2)/k_{rac} = 0.963\ k_{rac}^{-1} \tag{9}$$

from where both the radioracemization rate constant $k_{rac}$ and the half life of optical activity can be derived $T_{1/2rac}$.

3.2 - The amino acids stability to radiation and radioracemization: the results

The solid state irradiation of the amino acids causes the formation of free radicals in the molecular crystal followed by the decomposition of the pristine amino acid for instance by a deamination or a decarboxylation reaction (Sagstuen et al. 2004). Thus, the continuous radiolysis causes a steady decomposition but occurring in the solid state the radiolysis products are not so mobile and cannot escape easily, especially large molecular fragments remain trapped in the crystalline structure. With the DSC technique the measurement of the purity of a given crystalline compound is made through its melting enthalpy (Brown, 2001). Initially the melting enthalpy of a pristine, reference amino acid is measured and the onset and peak of the melting point determined as shown for example in Fig. 1 in the case of the amino acid alanine. Then, the same amino acid but after 3.2 MGy of radiation dose, is studied by DSC. As shown in Fig. 1, there is a significant reduction in the melting enthalpy of the irradiated sample and a shift of the melting point onset and peak. More in detail, pure alanine (Fig. 1) has a melting enthalpy of -1387.4 J/g and after the administration of 3.2 MGy it is reduced to -938.5 J/g.

The reduction of the melting enthalpy is due to the partial decomposition of the alanine into other products and hence its purity is reduced to $N_\gamma$ = 100 (-938.5/-1387.4) = 67.6%. The decomposition products do not participate to the melting enthalpy because do not form a crystalline structure and only the residual amino acid still possesses a crystalline structure, this justifies a reduction of the melting enthalpy. Similarly, the radiolysis products that remained
trapped in the crystalline structure of the amino acids cause a reduction of the melting point. Thus, in the case of alanine shown in Fig. 1, the onset and peak melting point of the pristine, sample occur at 291.0°C and at 293.5°C respectively but after the radiolysis at 3.2 MGy such peaks transitions appear shifted at lower temperatures 275.7°C and 282.4°C respectively. Of course there is a correlation between the reduction of the melting enthalpy and the shift toward lower temperature of the melting transition.
The other approach we have used to measure the degree of decomposition of all the amino acids was to measure their optical activity through the ORD measurement. The optical activity of any enantiomer is expressed through the Biot law (Djerassi, 1960; Jirgensons, 1973):

$$[\alpha]_\lambda = \alpha\ l^{-1} c^{-1} \tag{eq. 10}$$

where the specific optical rotation $[\alpha]_\lambda$ is a constant at a given wavelength and concentration c of the given optical active molecule; α is the rotation degree of the plane of polarized light after the passage through a polarimetric tube having an optical path length l. The value of $[\alpha]_\lambda$ changes only with the wavelength. Thus, the optical rotatory dispersion curve (ORD) is the value of $[\alpha]_\lambda$ as function of the wavelength as shown in Fig. 2 and 3.

Pristine, reference amino acid at a certain concentration in HCl 1M give an ORD curve or spectra like those shown in Fig. 2 in the case of l-alanine and l-leucine or in Fig. 3 in the case of l-tyrosine, l-phenylalanine and l-tryptophan. When the solid state irradiated amino acids (at 3.2 MGy) are dissolved in HCl 1M at the same concentration of the corresponding reference pristine amino acid, produce an ORD spectrum which appear shifted down in the vertical direction. Since l-alanine and l-leucine give a positive specific optical rotation, the corresponding irradiated samples show a down shift of the entire ORD curve toward the abscissa axis (see Fig. 2). On the other hand, since the pristine amino acids l-tyrosine, l-phenylalanine and l-tryptophan show a negative specific optical rotation, the corresponding irradiated samples at 3.2 MGy dissolved in HCl 1 M at the same concentration show an upper shift of the ORD curve again in the vertical direction (see Fig. 3). In both cases such shift is a consequence of the radioracemization of the sample. According to eq. 2, $R_\gamma$ is the residual amount of an amino acid enantiomer after 3.2 MGy radiation dose. The value of $R_\gamma$ was determined at all the wavelengths shown in Fig. 2 and 3 and averaged. The radiolysis of amino acids may cause a simple inversion of the chiral centre of the enantiomer irradiated. In such a case we are dealing with a true radioracemization which involves the isomerization of the l-enantiomer into the d-enantiomer. Such an isomerization will imply a reduction in the intensity of the specific optical rotation at any wavelength. Because of the high energies involved in the solid state γ radiolysis of the amino acids does not involve only the true radioracemization but also the so-called "false" or apparent radioracemization. The "false" or apparent radioracemization involves the radiolysis of the given enantiomer into other products with the loss of any chiral centre. In that case, the reduction of the $[\alpha]_\lambda$ value and the consequent shift of the ORD curve is not due to the transformation of one enantiomer into the other but to the decomposition of the enantiomer. The residual optical activity $R_\gamma$ will hence be due to the residual enantiomer that survived the radiolysis. There is also another possibility: the survival of the chiral centre in the radiolysis products. Such an event is rare but possible and, of course, will lead in any case to an alteration of the ORD curve in such a way that it will appear as an apparent racemization.

Table 1 summarizes all the experimental results regarding the solid state radiolysis of all proteinaceous amino acids studied both in terms of residual amino acid amount which survived the 3.2 MGy radiolysis measured by ORD spectroscopy ($R_\gamma$ %, eq. 2) and measured by DSC ($N_\gamma$ %, eq. 1). More in detail, Table 1 shows the proteinaceous amino acid ordered according to their radioracemization resistance from the least resistant to the most resistant. Threonine is the amino acid showing the lowest resistance to radioracemization and tyrosine is the most resistant. However, the most astonishing thing is that after 3.2 MGy radiation dose, a dose equivalent to $1.05 \times 10^9$ years of irradiation due to radionuclide decay, at least 62.8% of the original amount of l-threonine optical activity is preserved and the results are even more encouraging for all the other amino acids. Already some authors have reported about the surprising preservation of chirality to considerable radiation doses (Kminek and Bada, 2006; Cataldo 2007a; Cataldo et al 2008; Izumi et al. 2008) but the present study systematically examines the stability of all amino acids toward high energy radiation and the result is that practically in all cases they can survive an enormous radiation dose if administered in the solid state. Indeed Table 1 shows also a reasonable correlation between the amount of amino acids survived 3.2 MGy as measured from ORD spectroscopy ($R_\gamma$ %) in comparison to the amount of the amino acids survived the radiolysis as measured by DSC (Nγ %). As already discovered by Bonner and Lemmon (1978), as a general rule to a certain degree of the radiolytic decomposition of the amino acids does not correspond necessarily exactly the same degree of radioracemization.

| TABLE 1 - AMINO ACIDS ORDERED ACCORDING TO THE RADIORACEMIZATION RESISTANCE | | |
|---|---|---|
| | $R_\gamma$ % after $1.05 \times 10^9$ y | $N_\gamma$ % after $1.05 \times 10^9$ y |
| l-threonine | 62.8 | 66.6 |
| l-serine | 70.7 | 80.8 |
| l-cysteine | 71.4 | 63.5 |
| l-arginine | 75.3 | 56.6 |
| l-alanine | 75.8 | 67.6 |
| l-lysine | 77.0 | 54.6 |
| l-tryptophan | 79.1 | 92.2 |
| l-phenylalanine | 82.5 | 70.2 |
| l-methionine | 82.6 | 68.4 |
| l-leucine | 82.6 | 72.3 |
| l-proline | 86.8 | 83.5 |
| l-isoleucine | 91.4 | 70.5 |
| l-glutamine | 91.6 | 94.7 |
| l-glutamic acid | 92.7 | 84.5 |
| l-valine | 94.0 | 98.1 |
| l-aspartic acid | 94.2 | 95.5 |
| l-histidine | 96.3 | 83.4 |
| l-asparagine | 97.6 | 79.6 |
| l-cystine | 97.6 | 96.7 |
| l-tyrosine | 98.9 | 92.1 |

For example Table 1 shows that in terms of $N_\gamma$ the worse radiolysis resistance belongs to lysine which however is ranked 15$^{th}$ in the radioracemization list (left column of Table 1). On the other hand the best radiolytic stability in the solid state (in terms of $N_\gamma$) is offered by valine which also shows a considerable radioracemization resistance but is ranked only 6$^{th}$ in the radioracemization list. These differences between the amount of amino acid survived the radiolysis as measured by DSC and expressed as $N_\gamma$ and the amount of the same amino acids survived the radiolysis and measured by ORD spectroscopy and expressed by $R_\gamma$ can be explained by advocating the true and apparent radioracemization phenomena discussed earlier. In any case Fig. 4 provides a reasonable correlation between the $N_\gamma$ and $R_\gamma$ parameter so that the following equation has been derived:

$$N_\gamma = 0.864\, R_\gamma + 5.182 \qquad (eq.\ 11)$$

Although the $R^2$ value of 0.452 demonstrates a certain dispersion of the data.

Knowing the $N_\gamma$ and $R_\gamma$ values of each proteinaceous amino acid measured at 3.2 MGy which corresponds to $1.05 \times 10^9$ years of irradiation, we can extrapolate back these data to $4.6 \times 10^9$ years ago, i.e. the beginning of the Solar System. Table 2 provides a summary of such an extrapolation.

| TABLE 2 - RESIDUAL AMINO ACIDS AFTER EXTRAPOLATION TO 4.6 GIGAYEARS | | |
|---|---|---|
|  | $R_\gamma$ % after $4.6 \times 10^9$ y | $N_\gamma$ % after $4.6 \times 10^9$ y |
| l-threonine | 13.1 | 16.9 |
| l-serine | 21.9 | 39.3 |
| l-cysteine | 22.9 | 13.7 |
| l-arginine | 28.9 | 8.3 |
| l-alanine | 29.7 | 19.5 |
| l-lysine | 31.9 | 7.1 |
| l-tryptophan | 35.8 | 70.1 |
| l-phenylalanine | 43.1 | 21.3 |
| l-methionine | 43.3 | 19.0 |
| l-leucine | 43.3 | 24.2 |
| l-proline | 53.8 | 45.4 |
| l-isoleucine | 67.5 | 21.7 |
| l-glutamine | 68.1 | 78.8 |
| l-glutamic acid | 71.8 | 47.8 |
| l-valine | 76.3 | 91.9 |
| l-aspartic acid | 77.0 | 81.7 |
| l-histidine | 84.8 | 45.2 |
| l-asparagine | 89.9 | 36.8 |
| l-cystine | 89.9 | 86.3 |
| l-tyrosine | 95.3 | 69.8 |

Thus, in Table 2 the value of $R_\gamma = 13.1\%$ means that only such fraction of the original optical activity due to threonine survived from the beginning of the Solar System till the present time and, similarly only 16.9% of the threonine present at the beginning of the Solar System survived the radiolysis till the present day. However there are amino acids with considerable radiation resistance whose amount should be approximately the same as in the origin of the Solar System. It is the case of valine, or other amino acids like aspartic acid and glutamine which are expected to be today just 4/5 of the original amounts present at the beginning of the formation of the Solar System. The preservation of chirality is even more impressive: after 14 MGy administered in the solid state, the amino acids glutamic acid, valine, aspartic acid, histidine, asparagine, cystine and tyrosine still display >70% of the original chirality they had $4.6 \times 10^9$ years ago.

Based on these data it is not at all a surprise that amino acids have been found in meteorites and in measureable chiral excess. Furthermore, this study supports experimentally the theory (Crovisier and Encrenaz, 2000) that amino acids were formed in the interstellar medium and in chiral excess and then have been incorporated in comets and asteroids at the epoch of the Solar System formation.

Using eq. 5 and eq. 8 we have determined the radiolysis rate constant of each amino acid $k_{dsc}$ and the radioracemization rate constant $k_{rac}$ in year$^{-1}$ as reported in Table 3 respectively. As expected from the previous discussion on the $N_\gamma$ and $R_\gamma$ values, the $k_{dsc}$ and $k_{rac}$ have the same order of magnitude ranging from $10^{-10}$ y$^{-1}$ for the less radiation resistant amino acids to $10^{-11}$ y$^{-1}$ for the most radiation resistant. Similarly also the half life of each amino acid has been determined using the eq. 6 and 9. Again the $T_{1/2dsc}$ and $T_{1/2rac}$ have the same order of magnitude in the range between $10^9$ and $10^{10}$ year as reported in Table 3.

The validity of our experimental approach can be checked y comparing the $k_{dsc}$ and $T_{1/2dsc}$ results derived by calorimetric measurements with those of Kminek and Bada (2006). These authors have irradiated alanine in the solid state with γ radiation and have measured the $N_\gamma$ by HPLC analysis instead of the DSC used by us. For alanine they have found a radiolysis rate constant k = 3.43x10$^{-10}$ y$^{-1}$ in good agreement with the value found in the present work (Table 3) k = 3.56x 10$^{-10}$ y$^{-1}$. Kminek and Bada (2006) report also k = 5.26x10$^{-10}$ y$^{-1}$ for glutamic acid and k = 4.78x 10$^{-10}$ for aspartic acid, while our results, as reported in Table 3 for these two amino acids, are respectively k = 1.60x 10$^{-10}$ y$^{-1}$ and k = 0.438x10$^{-10}$ y$^{-1}$, in fair agreement.

| TABLE 3 - RADIORACEMIZATION AND RADIOLYSIS RATE CONSTANTS AND RELATIVE HALF LIFES | | | | |
|---|---|---|---|---|
| | $k_{rac}$ (x10$^{-10}$ y$^{-1}$) | $k_{dsc}$ (x10$^{-10}$ y$^{-1}$) | $T_{1/2rac}$ (x10$^9$ y) | $T_{1/2dsc}$ (x10$^9$ y) |
| l-threonine | 4.43 | 3.87 | 1.57 | 1.79 |
| l-serine | 3.30 | 2.03 | 2.10 | 3.42 |
| l-cysteine | 3.21 | 4.32 | 2.16 | 1.60 |
| l-arginine | 2.70 | 5.42 | 2.57 | 1.28 |
| l-alanine | 2.64 | 3.56 | 2.63 | 1.95 |
| l-lysine | 2.49 | 5.76 | 2.79 | 1.20 |
| l-tryptophan | 2.23 | 0.773 | 3.11 | 8.97 |
| l-phenylalanine | 1.83 | 3.37 | 3.79 | 2.06 |
| l-methionine | 1.82 | 3.61 | 3.81 | 1.92 |
| l-leucine | 1.82 | 3.09 | 3.81 | 2.25 |
| l-proline | 1.35 | 1.72 | 5.15 | 4.04 |
| l-isoleucine | 0.856 | 3.33 | 8.10 | 2.08 |
| l-glutamine | 0.835 | 0.518 | 8.30 | 13.4 |
| l-glutamic acid | 0.721 | 1.60 | 9.61 | 4.33 |
| l-valine | 0.589 | 0.183 | 11.8 | 38.0 |
| l-aspartic acid | 0.569 | 0.438 | 12.2 | 15.8 |
| l-histidine | 0.359 | 1.73 | 19.3 | 4.01 |
| l-asparagine | 0.231 | 2.17 | 30.0 | 3.19 |
| l-cystine | 0.231 | 0.319 | 30.0 | 21.7 |
| l-tyrosine | 0.105 | 0.783 | 65.9 | 8.85 |

## 4. Conclusions

The 20 proteinaceous amino acids studied can survive a radiation dose of 14 MGy, which is the dose estimated to be delivered along the life of the Solar System to all organic molecules present in comets or asteroids at a depth of 20 m or more. The dose of 14 MGy is due to the decay of all major radionuclides in $4.6 \times 10^9$ years. At a depth >20 m, the contribution from cosmic rays to this total dose is negligible because of the shielding.

Our study shows that eachno acid has an individual response to high energy radiation but it is evident that all the proteinaceous amino acids can survive in relatively large amount to a radiation dose equivalent to that administered by the radionuclide decay from the beginning of the history of the Solar System to the present. Not only all the amino acids can survive to 14 MGy but also their chirality can be partially preserved even after such radiation dose.
Therefore, based on these results it is not at all a surprise that amino acids have been found in meteorites and in measureable chiral excess. Furthermore, this study supports experimentally the theory (Crovisier and Encrenaz, 2000) that amino acids were present in the interstellar material that conformes the protostelar nebula and in chiral excess. Then they were incorporated in comets and asteroids at the epoch of the Solar System formation. With the experimental data available we can even predict the concentration of the amino acids in comets and in asteroids at the beginning of the Solar System and also we can extrapolate to the original chiral excess $4.6 \times 10^9$ years ago.

## Acknowledgements

The present research work has been supported by grant AYA2007-64748 of the Spanish Ministerio de Ciencia e Innovacion.

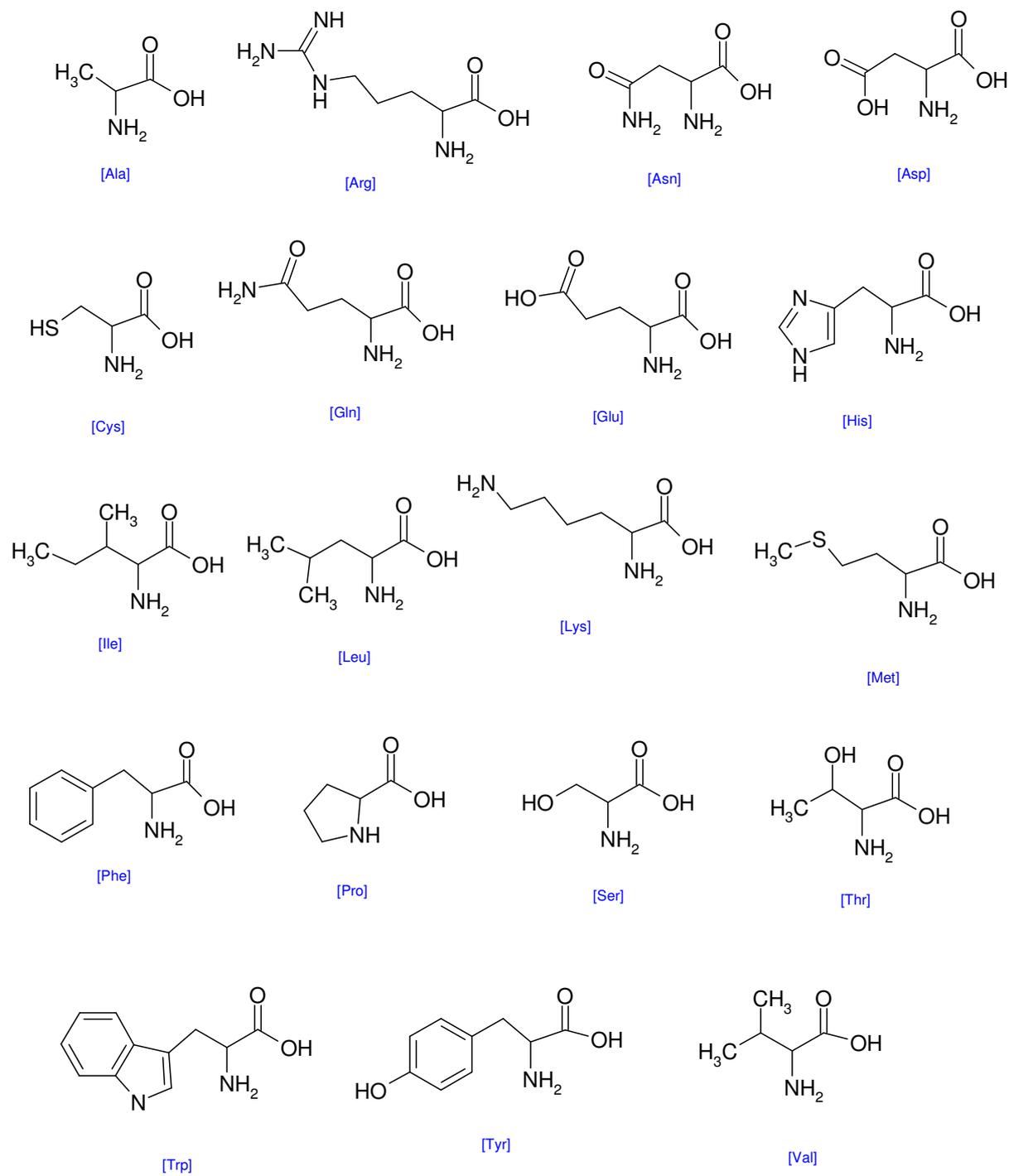

SCHEME 1

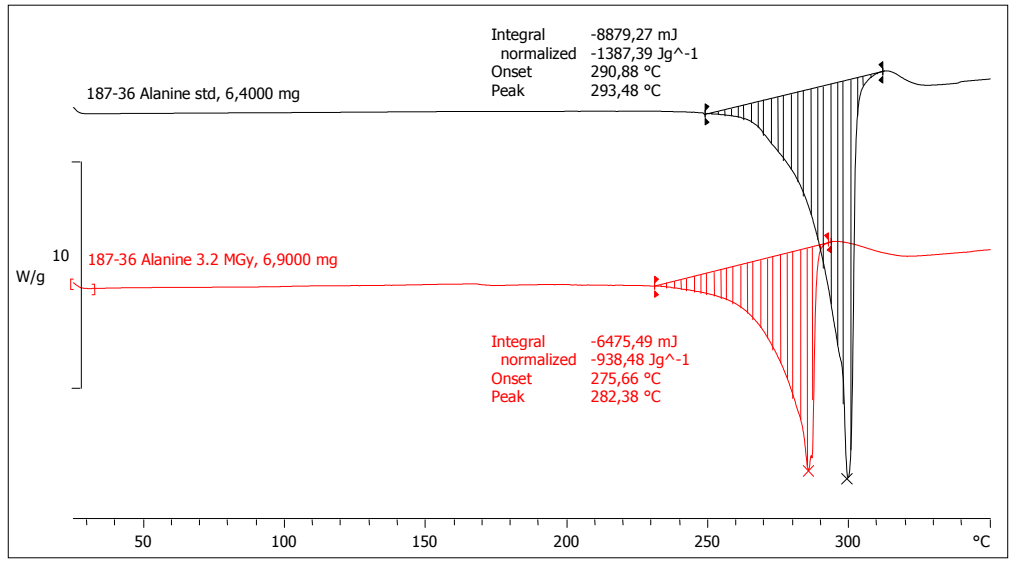

Fig. 1 – DSC of alanine in Al crucible heating rate 10°C/min. The upper trace is due to standard pristine alanine and the lower trace is due to the 3.2 MGy treated sample.

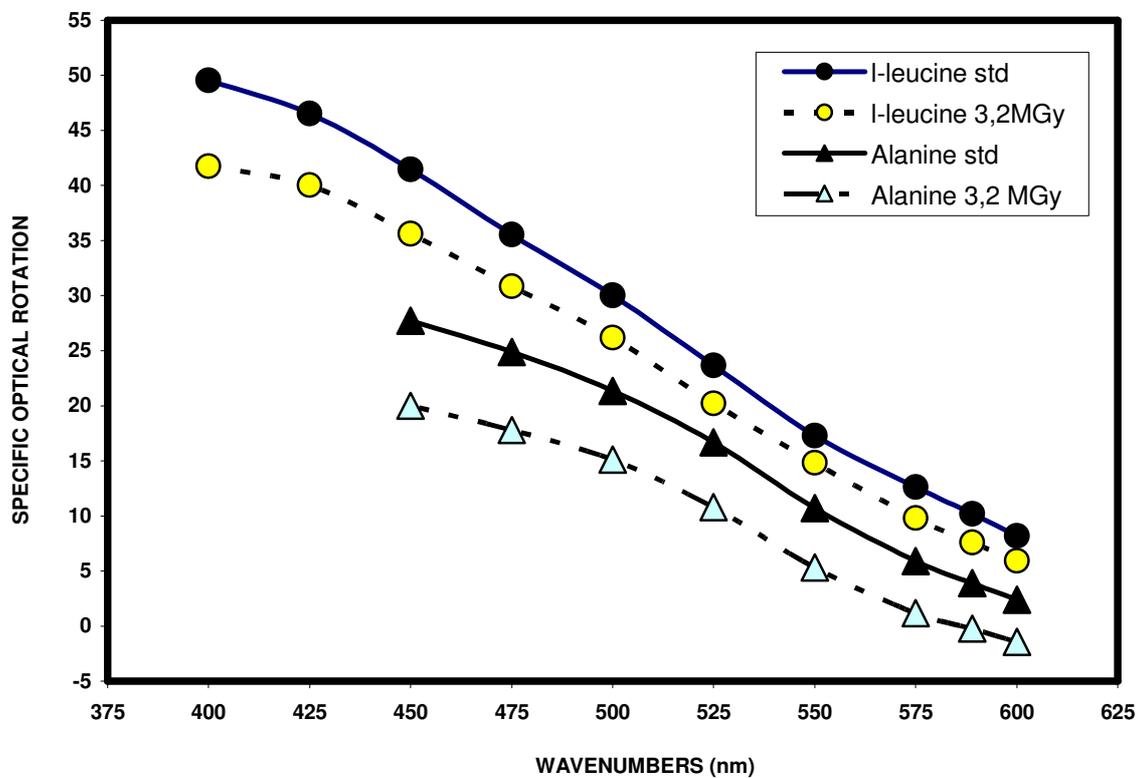

Fig. 2 – Optical rotatory dispersion (ORD) of leucine and alanine before and after the solid state radiolysis at 3.2 MGy. The extent of the radioracemization can be appreciated by the shift of the ORD curve toward the abscissa after the radiolysis. Concentration in HCl 1M (pristine and irradiated): alanine 4.1 mg/ml; leucine 5.2 mg/ml.

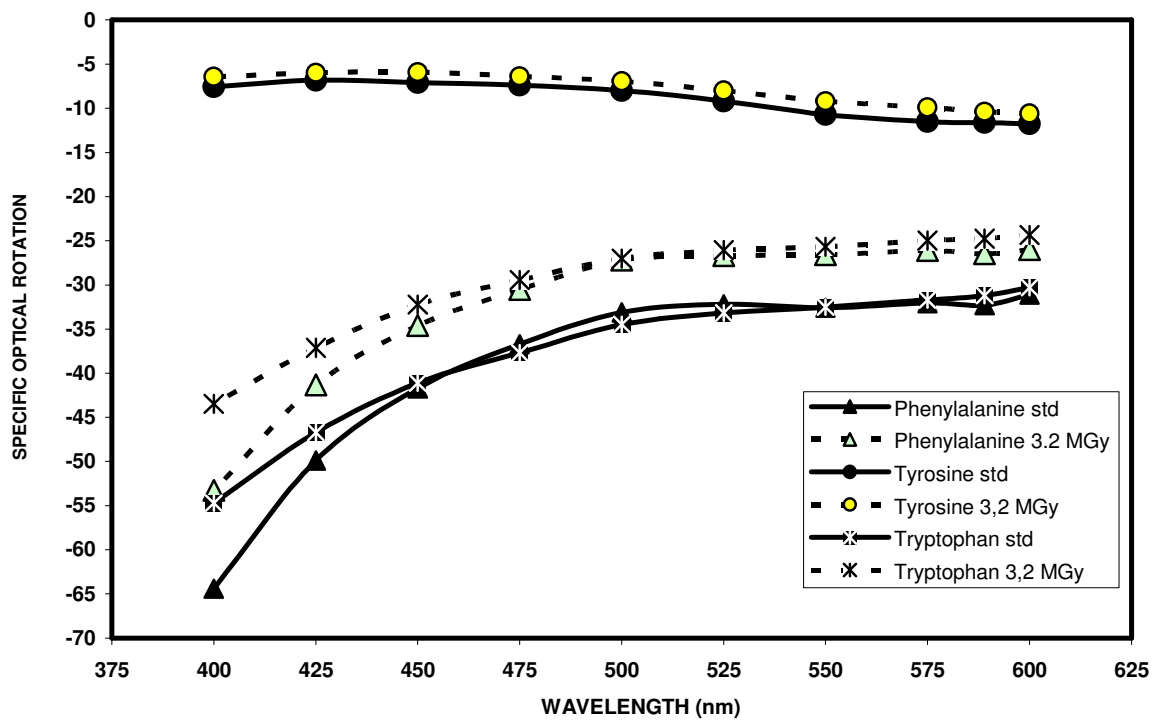

Fig. 3 – Optical rotatory dispersion (ORD) of phenylalanine, tyrosine and tryptophan before and after the solid state radiolysis at 3.2 MGy. The extent of the radioracemization can be appreciated by the shift of the ORD curve toward the zero axis after the radiolysis. The radioracemization is minimal in the case of tyrosine and appreciable. Concentration in HCl 1M (pristine and irradiated): phenylalanine 5.0 mg/ml; tyrosine 10.0 mg/ml; tryptophan 8.0 mg/ml.

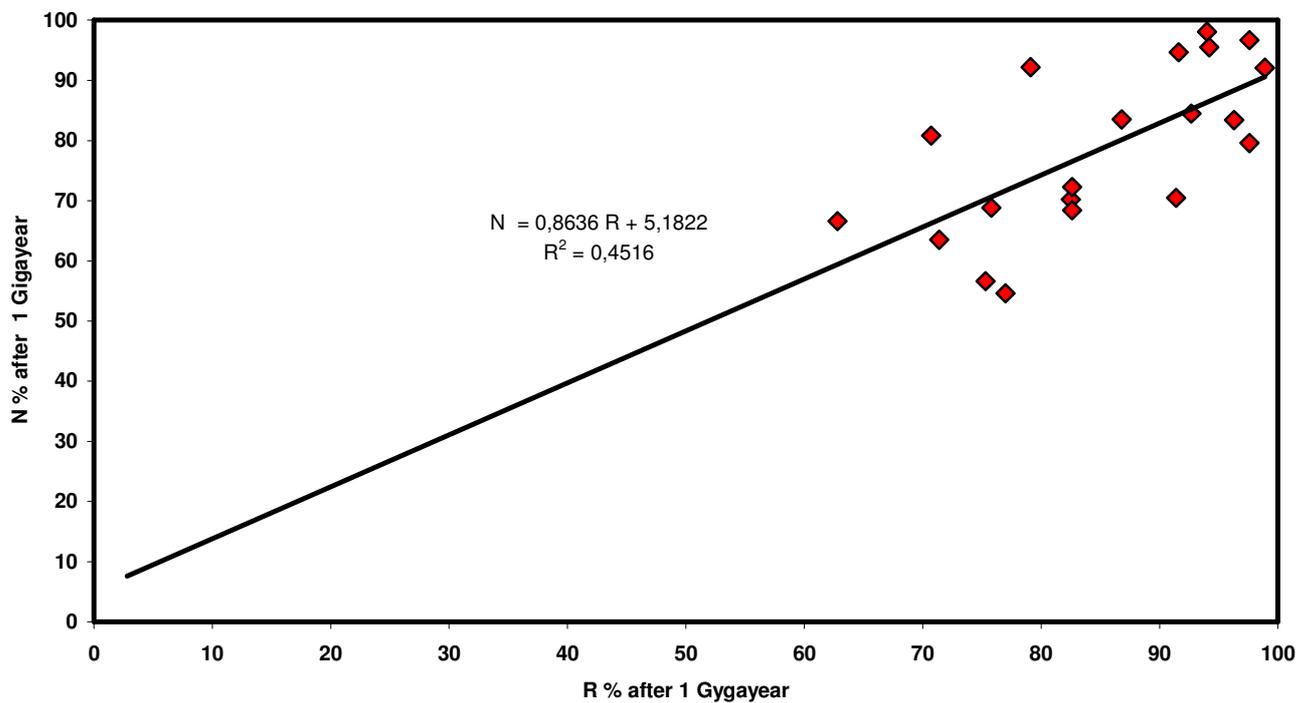

Fig. 4 – Correlation between the residual amount of amino acid survived the radiolysis equivalent to 1 Gigayear as measured by DSC (N %) and amount measured by ORD (R %). There is correlation but the correlation index $R^2$ is not particularly high, indicating a certain dispersion of the two set of data.